\documentclass[superscriptaddress]{revtex4}
\usepackage{graphicx}
\usepackage{times}

\begin{document}
\title{OPEN QUESTIONS IN CMR MANGANITES, \\
RELEVANCE OF CLUSTERED STATES, AND
ANALOGIES WITH OTHER COMPOUNDS}

\author{ELBIO DAGOTTO}
\affiliation{National High Magnetic Field Lab and Department of Physics,
Florida State University, Tallahassee, FL 32306, USA ~~~ dagotto@magnet.fsu.edu}

\date{\today}

\begin{abstract}
This is an informal paper that contains a list of ``things we know'' and ``things
we do not know'' in manganites. It is adapted from the conclusions chapter of
a recent book by the author, {\it Nanoscale Phase Separation and Colossal 
Magnetoresistance. The Physics of Manganites and Related Compounds}, 
Springer-Verlag, Berlin, November 2002. The main new result
of recent manganite investigations is the discovery of tendencies toward
inhomogeneous states, both in experiments and in simulations of models. 
The colossal magnetoresistance effect appears to be closely linked to these 
mixed-phase tendencies, although 
considerably more work is needed to fully confirm these ideas.
The paper also includes
information on cuprates, diluted magnetic semiconductors, relaxor ferroelectrics, 
cobaltites, and
organic and heavy fermion superconductors. These materials potentially share 
some common phenomenology with the manganites, such as a temperature scale $T^*$ 
 above the ordering temperature where anomalous behavior starts. Many of these
materials also present
low-temperature phase competition.
The possibility of colossal-like effects in compounds that do not involve
ferromagnets is briefly discussed. Overall, it is concluded that inhomogeneous
``clustered'' states should be 
considered a new paradigm
in condensed matter physics, since their presence appears to be far more common
than previously anticipated.

\end{abstract}

\maketitle

\section{\bf Introduction}

In this paper -- not intended for publication in a formal journal --
I present a list of ``things we know'' and ``things we do not
know'' about manganites and related compounds. This text is adapted
from the last chapter of a book I recently finished on 
manganites \cite{book}, after receiving the suggestion by some
colleagues of making the ``open questions''
discussion available to a wider readership. The presentation is
very informal to keep the discussion fluid, and it is intended for
researchers with some background in manganites (other materials, such
as cuprates, are also addressed). Only a handful of
references are included here for simplicity. However, close to 1,000
citations
can be found in the original source \cite{book} as well as the detailed
justification of many of the matter-of-fact statements expressed
here, particularly in the ``things we know'' section. 
Many reviews are also available \cite{review} with plenty of
references: this paper is not a review article but an informal
discussion, with many comments and sketchy ideas.
Also, some 
items reflect the personal opinion of the author and may be debatable.
In addition, this paper discusses
results for other families of compounds which exhibit similarities
with manganites. It is conceivable
that the knowledge accumulated in Mn-oxides may be applicable to the
famous high-T$_c$ cuprates, as well as other materials 
discussed here and in other chapters of \cite{book}.
The stability of ``clustered'' states -- often called 
phase-separated or mixed-phase states -- 
appears to be an intriguing property 
of a variety of compounds, and ``colossal effects'' 
should be a phenomenon far more common than
previously believed. There is plenty of
work ahead, and surprises waiting to be unveiled.

\section{\bf 
Facts About Manganites Believed to be Understood (``things we know'')}

After the huge effort carried out in recent years in the study of
manganites, involving both experimental and theoretical
investigations, much progress has been achieved, described in the
bulk of \cite{book} and in \cite{review}. Here, 
a brief list of established results is presented.
In the next section, the detail of the discussion will increase
when issues that remain to be understood are addressed.

\vskip 0.5cm

{\bf 1.} At hole density $x$=0, as in $\rm LaMnO_3$, the ground state  
has staggered orbital order (concept first discussed by Khomskii and Kugel). 
The spins form
an A-type antiferromagnet (antiparallel spins along one direction, parallel along the
other two), although 
very recently more exotic spin arrangements at $x$=0 have been proposed for
some manganites \cite{newphases}. 
Several other doped Mn-oxides have orbital order (OO) as well.
This orbital ordering appears to be triggered by a large electron -- Jahn-Teller 
phonon coupling, as discussed by Kanamori and
more recently in Mn-oxides by Millis and collaborators.
The orbital degree of freedom plays a role as important as
the spin, charge, and lattice degrees of freedom. This leads to 
a complex phase diagram, with many competing phases, a characteristic
of most correlated electron systems. I believe it is time to fully bring
manganites within the umbrella of ``correlated electrons'', since
the subject's complexity goes well-beyond standard issues of plain ``magnetism''.

\vskip 0.3cm
{\bf 2.} At $x$=1, fully hole doped, the spin arrangement is of G-type (antiparallel
spins in the three directions).

\vskip 0.3cm
{\bf 3.} There are many manganites with a metallic ferromagnetic (FM) ground state at
intermediate densities. The Zener mechanism --for historical reasons
widely known as ``double exchange''-- appears to capture the essence of the 
tendency to ferromagnetism. This FM metallic state
is a poor metal, and it is only one of the several possible states of  
manganites. Some authors believe that 
orbital correlations may play a role in this ferromagnet. It is also
unclear to what extent oxygens should be explicitly 
included in the electronic sectors of manganite models, or whether they can
be ``integrated out'' as usually assumed in theoretical studies of Mn-oxides.
This also occurs in $t$-$J$ models for cuprates that invoke Zhang-Rice singlets and
only focus on the Cu sites.
Thus, there
is still some discussion about the fine details of this state, but the
essential features of the FM phase appear to be qualitatively understood.

\vskip 0.3cm
{\bf 4.} 
At other hole densities, such as $x$=0.5, charge/orbital/spin ordered states can
be stabilized, leading to a
phase diagram interestingly asymmetric with respect to half doping.
The existence of these half-doped charge-ordered states
has been established in Monte Carlo
simulations, as well as in mean-field studies, 
in spite of their complicated structure. This gives hope that the
theoretical studies are on the right track. The charge ordered
(CO) states strongly
compete with the metallic ferromagnetic state, both in real experiments and
theoretical studies. 
This is a key observation leading to
potential explanations of the CMR,
since this phenomenon appears to occur when two phases --metal and 
insulator-- are in competition. It happens at the boundaries between
FM metal and paramagnetic insulator or FM metal and CO/OO/AF states.
This line of thinking -- CMR as caused by phase competition -- is
relatively new but currently dominates the literature. We simply cannot
understand the CMR effect from the ferromagnetic metallic state alone.

\vskip 0.3cm
{\bf 5.} The existence of intrinsic inhomogeneities 
in single crystals has been established experimentally (the list of
experimental groups involved is too long to be provided here, 
please see \cite{book,review}). 
This can occur within the ordered phases, or above the ordering 
temperatures, and even in the high hole-doping region (see work by
Neumeier and collaborators). 
Charge-ordered {\it nanoclusters} above the Curie temperatures have been 
found in many low- and
intermediate-bandwidth manganites. In the interesting CMR regime
this behavior is correlated 
with the resistivity, according to neutron scattering experiments
carried out by groups in Japan, Oak Ridge, Maryland, Brookhaven, Rutgers,
and other locations,
that have studied hidden structures near Bragg peaks (diffuse scattering).
These nano-clusters are believed to be crucial for the occurrence
of CMR \cite{newnano}.

\vskip 0.3cm
{\bf 6.} Related with the previous item,
percolative tendencies in manganites have been unveiled
experimentally and in theoretical studies. But sharp
first-order metal-insulator transitions
have also been reported, again both in theory and
experiments. In some compounds, a mixture of these behaviors
--percolation and first-order features-- have been found (for
more information, see Section VI below). These different behaviors
may depend on the strength of the inevitable disorder in the samples.

Some experiments -- notably those by Cheong's and Mydosh's groups --
have revealed the presence of submicrometer-size regions, to be
contrasted with the more common occurrence of nanometer-size clusters 
(see above). This is
a remarkable effect that deserves to be analyzed in detail. 
It is unclear to what extent the submicrometer clusters are truly
needed to understand the CMR and whether they are present in single crystals. 

\vskip 0.3cm
{\bf 7.}  The presence of phase separation tendencies in theoretical models
has been unveiled and confirmed, after the work of Yunoki, Moreo and 
collaborators in 1998 using Monte Carlo simulations of realistic models,
followed by a large number of other studies that reached similar 
conclusions (see work by Arovas, Guinea, Khomskii, Kagan, and others). 
Phase separation can be caused by at least two
tendencies: (1) {\it Electronic phase separation}, where competing states 
have different hole
densities and $1/r$ Coulomb effects lead to nanocluster coexistence. 
This is similar in spirit to the phase separation much discussed
in cuprates (see Kivelson, Tranquada, Zaanen, Di Castro and others).
(2) Disorder-driven phase
separation near first-order transitions (see \cite{moreo}), 
with competing states of equal density that leads to
large coexisting clusters.

\vskip 0.3cm
{\bf 8.} Theories based on states composed of ferromagnetic small
islands with randomly oriented magnetizations appear to capture the
essence of the CMR phenomenon \cite{book,review,burgy}. 
Electrons with spin, say, up are mobile 
 within ferro clusters with magnetization up. However,
they cannot travel through magnetization-down clusters, which
effectively act as insulators to the spin-up electron carriers.
However, the large effective spin of these
clusters can be easily rotated upon the application of very modest
fields, much smaller than the natural scales in the problem, leading
to a metallic state \cite{burgy}. Note that in these simulations
there was no need to carry out studies using submicrometer size 
clusters, since nanoclusters were sufficient.

\vskip 0.3cm
{\bf 9.} Theoretical studies have predicted that the metal and
insulator phases of manganites are separated by first-order transitions
leading to bicritical or tricritical behavior  in the clean limit
(see \cite{book,review,burgy}). Experiments by Tomioka and Tokura
have recently reported an example with such a behavior. 
When quenched disorder is sufficiently
strong, a ``window'' with disorder characteristics opens in between
the two phases as shown in the figure below. The group of Ibarra has 
reported results for a material that indeed has a glassy state between
the FM metal and the AF insulator. The particular case
of a quantum critical point is only obtained in this context by
fine-tuning of parameters
(although
multicritical phenomena may still be of relevance in this context
as proposed by Nagaosa and collaborators).

\vskip 0.3cm
{\bf 10.} Related to the previous item,
theoretical and experimental studies have unveiled the
existence of a new temperature scale $T^*$ where clusters start
forming well above the Curie temperature. Independent studies by
Burgy et al. and Salamon et al. have characterized this critical
temperature as a Griffiths temperature. The Griffiths effects appear
to be larger than usual due to phase competition.

\begin{figure}[h]
\includegraphics[width=.85\textwidth]{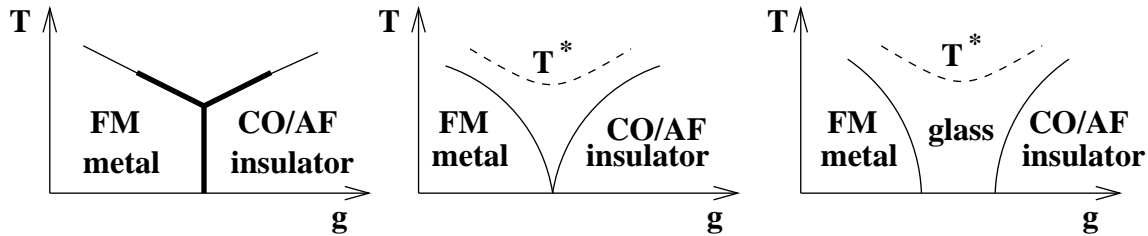}
\caption[]{
({\it Left}) Generic phase diagram of two competing phases in the absence of
quenched-disorder (or when this disorder is very weak). 
Thick (thin)  lines denote
first (second) order transitions. Shown is a tricritical case, but it could be
bicritical or tetracritical as well. 
$g$ is some parameter needed to change from one phase
to the other. ({\it Middle}) With increasing disorder the temperature range
of first-order transitions separating the ordered states is reduced, and
eventually for a {\it fine-tuned} value of the disorder the resulting
phase diagram contains a quantum critical point. In
this context this should be a rare occurrence. ({\it Right}) In the limit of
substantial disorder, a window opens between the ordered phases. The state
in between has glassy characteristics and it is composed of coexisting clusters
of both phases. The size of the coexisting islands can be regulated by 
disorder and by the proximity to the original first-order transition.
For more details see \cite{book,review,burgy}. 
The $T^*$ discussed in the previous item
-- remnant of the clean-limit transition -- is also shown.
\label{sketch.eps}
}
\end{figure}

\section{\bf Can theories that do not address phase separation work to understand
the CMR effect?}

The enormous experimental effort on Mn-oxides has already provided 
sufficient results
to decide whether or not some of the theories proposed in recent years 
realistically explain the unusual magnetotransport properties
of these compounds. Indeed, some of the
early proposals for theories of CMR manganites have already been shown 
to be incomplete
and the current leading effort centers around inhomogeneities and 
first-order metal-insulator transitions.
The detail is the following:

\begin{figure}[h]
\includegraphics[width=.60\textwidth]{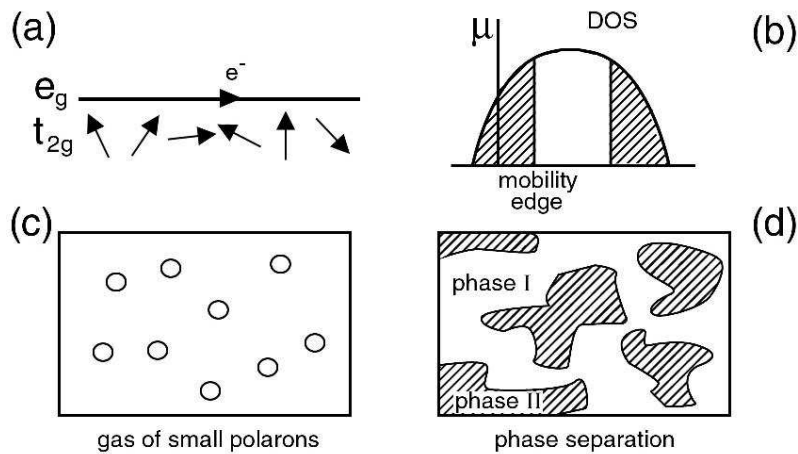}
\caption[]{
Schematic representation of theories for manganites. (a) is a simple
``double exchange'' scenario, without phase competition. 
(b) relies on Anderson localization as the origin of the insulating state
that competes with ferromagnetism. (c) is based on
a gas of polarons above the Curie temperature 
$T_{\rm C}$, also without phase competition. In (d), a phase-separated state above
the ordering temperatures is sketched. Details can be found in the text.
\label{theories.eps}
}
\end{figure}

\vskip 0.3cm
{\bf 1.} In theories sometimes 
referred to as ``double exchange'', electron hopping
above the Curie temperature $T_{\rm C}$ is simply described by a renormalized
hopping ``$t \langle cos \theta/2 \rangle$''. 
These theories are based on the movement of electrons
in a disordered spin-localized background
(see Fig.\ref{theories.eps} (a), for a crude sketch), without invoking other
phases. However, quantitative investigations have shown that this approach
 does not appear to be sufficient
to produce neither
an insulating state nor a CMR phenomenon, although this simple idea 
may be suitable for large
bandwidth manganites, such as $x$=0.4 LaSrMnO.

\vskip 0.3cm
{\bf 2} Some theories rely on Anderson localization
to explain the insulating state above $T_{\rm C}$. However,
the amount of disorder needed to achieve localization 
at the densities of relevance is very large
(at least in a simple three-dimensional Anderson model).
Computational studies \cite{slevin} 
locate the critical value
at $W_c$$\sim$16 (in units of the hopping), assuming a uniform 
box distribution of random on-site energies
[-$W/2$,+$W/2$], 
and with the Fermi energy at the band
center. Perhaps this large disorder strength crudely mimics 
the influence of large electron-phonon couplings, strong Coulomb
correlations, nanoclusters, strain,
and quenched disorder present in the
real materials. But, even in this case, it is difficult to
explain the density-of-states pseudogap found in photoemission
experiments by Dessau's group. Anderson
localization does not produce such a pseudogap 
(Fig.\ref{theories.eps} (b)). In addition,
experimentally it is clear 
that the CMR originates in the competition
between phases, typically FM-metallic and AF/CO-insulating, but  
this fundamental effect is not included in simple 
Anderson localization scenarios, where ordering
and phase competition are absent. For these reasons, I believe 
Anderson localization does not seem to be the best approach to explain
manganite physics.

\vskip 0.3cm

{\bf 3.} Some theories are based on a picture in which the 
paramagnetic insulating
state is made out of a gas of small and heavy polarons
(Fig.\ref{theories.eps} (c)).
These theories
do not address neutron experiments reporting CE 
charge-ordered clusters above $T_{\rm C}$, correlated 
to the resistivity, nor the many
indications of inhomogeneities. A polaron gas may be
a good description at much higher temperatures, well above room 
temperature, but such a state does not appear sufficient
in the region for CMR, close to the Curie temperature. 
The charge-ordered small
clusters found experimentally above $T_{\rm C}$ 
have properties corresponding to phases
that are stable at low temperatures, 
such as the CE-phase. These complex clusters certainly cannot 
be considered to be mere polarons. They are more like ``correlated
polarons'', as some researchers in this field prefer to call the
charge-ordered islands.

\vskip 0.5cm

Theories based on microscopic phase separation 
(Fig.\ref{theories.eps} (d)) appear to provide 
a more realistic starting point 
to manganites since they are compatible with dozens of experiments. However,
as shown below, there is plenty of room for improvement as well!

\section{\bf Some of the Open Issues in Mn-oxides (``things we do not know'')}

\subsection{Potentially Important Experiments (in random order)}

\vskip 0.2cm
\noindent $\bullet$ {\bf The evidence for charge-ordered 
nanoclusters above $T_{\rm C}$ should be further confirmed}. It is 
important
that a variety of techniques reach the same conclusions regarding the
presence of nanoclusters above $T_{\rm C}$, definitely ruling out
an homogeneous state as the cause of the CMR effect.
For instance, it would be important to find out the role played by
nanoclusters on
optical conductivity results, which thus far has been described mainly
as consisting of ``small polarons'' (Noh and collaborators). 
How do the nanoclusters manifest themselves
in the optical spectra? Recent results at $x$$>$0.5 by Noh's group
much contribute to this issue, since a pseudogap was observed compatible
with photoemission results.

\vskip 0.3cm

\begin{figure}[h]
\includegraphics[width=.32\textwidth]{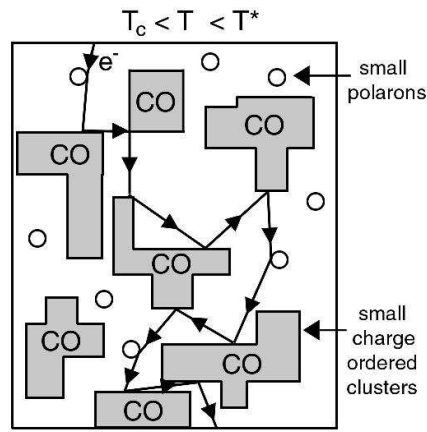}
\caption[]{
Crude cartoon representing the proposed
state in the regime between $T_{\rm C}$ and $T^*$.
Small polarons are represented by circles, and electrons traveling across
the sample (thick line) 
may not scatter much from them. The grey areas are the CO clusters
where scattering is more severe. The FM regions, expected to be 
present in this state according to some experiments, are not shown.
They can also contribute substantially to scattering, since their
orientations are random.
\label{cartoon.eps}
}
\end{figure}

\noindent $\bullet$ {\bf The existence of the predicted 
new temperature scale $T^*$ above the Curie temperature 
should be further investigated}.
(i) Thermal expansion, magnetic susceptibility, X-rays, neutron scattering,
and other techniques have already provided results supporting the
existence of a new scale $T^*$, where clusters start forming 
upon cooling. In fact, very early in manganite investigations, the group of
Ibarra at Zaragoza
reported the existence of such a scale, in  agreement with more
recent theoretical and experimental developments. 
This scale should manifest itself
even in the d.c. resistivity, as it does in the high temperature 
superconductors at the analog $T^*$ ``pseudogap temperature''. Are there
anomalies in $\rho_{dc}$ vs. temperature in Mn-oxides as well? Are the $T^*$
scales in cuprates and manganites indicative of similar physics?
(ii) In addition, the specific heat should systematically 
show the existence of structure at $T^*$ due to the development of 
short-range order (at $T^*$, even a glassy phase transition may occur, as
recently proposed by Argyriou et al.).
(iii) The dependence of $T^*$ with doping and tolerance factors should be
analyzed systematically. Theoretical studies \cite{burgy}
suggest that the tolerance factor may not change $T^*$
substantially, although it affects the ordering temperatures dramatically.
Is there experimental support for this prediction?
(iv) Is the crude picture of the
state between $T_{\rm C}$ and $T^*$ shown in
Fig.\ref{cartoon.eps} qualitatively correct?

\vskip 0.3cm

\noindent $\bullet$ {\bf X-rays and neutron-scattering studies 
of  $\rm {(La_{1-{\it y}} Pr_{\it y})_{1-{\it x}} Ca_{\it x} Mn O_3}$
(LPCMO)
are needed
to analyze the evolution of charge-ordered nanoclusters}. 
There is considerable evidence that LCMO $x$=0.3
at temperatures above $T_{\rm C}$ presents charge ordered nanoclusters, 
correlated with the behavior of the resistivity, as already discussed. It is 
important to track the intensity and location in momentum space of
the peaks associated with charge ordering as La is replaced by Pr. This
replacement
enhances charge ordering tendencies, as discussed by Cheong and 
collaborators. It is also important to notice that 
the substitution of 
Ca by Sr decreases that tendency. 
Is there a smooth evolution from LCMO to the regime found by Cheong's group in
LPCMO with phase separation at low temperatures, as well
as from LCMO to the more ``double exchange'' homogeneous
regime of LSMO?

\vskip 0.3cm
\noindent $\bullet$ 
{\bf Are the mesoscopic clusters found in LPCMO using electron 
diffraction representative of the bulk? Are there other compounds with
the same behavior?} 
The evidence for nanocluster formation appears robust
in manganites. However, the evidence for larger
structures of the mesoscopic kind is based on a smaller number of 
experiments: electron diffraction
and STM techniques (see Sec. II).
Are these results 
representative of single-crystal behavior? To what extent should one consider
two kinds of phase separation, i.e., nanoscale and microscale? Can one evolve
smoothly from one to the other as the Curie temperature decreases?

\vskip 0.3cm
\noindent $\bullet$ {\bf What is the nature of the 
ferromagnetic-insulating phases that appear in some phase diagrams?} 
Is this phase truly qualitatively
different from the ferromagnetic-metallic phase, or are they 
very similar microscopically? 
In other words, is there a spontaneously broken symmetry in the
ferro-insulator state? 
Does charge ordering and/or orbital ordering exist there?
Theoretical studies by Yunoki, Hotta, and others have shown the presence
of many novel spin ferromagnetic phases, with or without charge and orbital
order \cite{newphase}. 
There is no reason why these phases could not be stabilized in
experiments. In fact, a new ``E-phase'' of manganites at $x$=0
has  recently been discussed  \cite{newphases}. 
This is an exciting new area of
investigations and plenty of phases found in simulations could be 
observed experimentally, as recently exemplified by the ferromagnetic
charge-ordered phase discussed by Loudon and collaborators 
(see \cite{newphases}), which had been 
predicted by Yunoki {\it et al.} \cite{newphase}.

\vskip 0.3cm
\noindent $\bullet$ {\bf Atomic resolution STM experiments 
should be performed, at many
hole densities}. The very recent atomic-resolution STM results for 
BiCaMnO by Renner et al. are very important for the
clarification of the nature of manganites, and for the explicit visual
confirmation of
phase separation ideas. Extending experiments of this
variety to other hole densities, particularly those 
where the system becomes ferromagnetic metallic at low temperatures,
is of much importance. In the area of cuprates, high resolution STM results
by Davis, Pan and collaborators have made a tremendous impact, and hopefully
similar experiments can be carried out in manganites.

\begin{figure}[h]
\includegraphics[width=.5\textwidth]{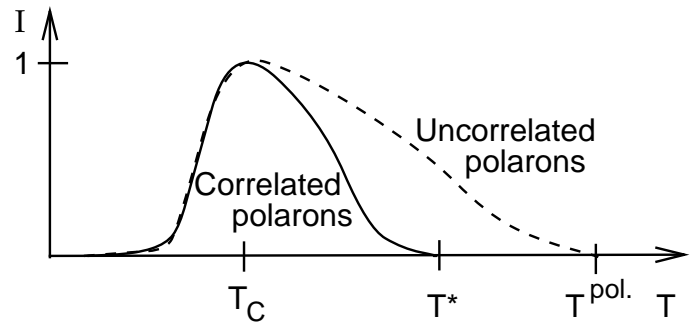}
\caption[]{
Schematic representation of an idealized diffuse neutron-scattering experiment.
Intensities $I$ in the vertical axis correspond to momenta
associated with (1) correlated polarons (a.k.a. charge-ordered clusters), 
and (2) uncorrelated polarons. Two
proposed temperature scales, $T^*$ and $T^{\rm pol}$, are shown.
The results are normalized to the same intensity at the Curie temperature
$T_{\rm C}$. For a discussion, see text and \cite{book}. These are theoretical
proposals, that need further experimental confirmation.
\label{correlated.polarons.eps}
}
\end{figure}

\vskip 0.3cm
\noindent $\bullet$ {\bf The regime of temperatures well above room temperature
must be carefully explored, even beyond $T^*$}. 
At $T$$>$$T^*$, the CO clusters are no longer formed but a gas of
polarons may still be present. Perhaps this leads to another temperature
scale, denoted by $T^{\rm pol}$ in Fig.\ref{correlated.polarons.eps},
where individual polarons start forming? Is the system metallic upon
further increasing the temperature? We may have interesting physics
even at temperatures close to 1,000 K in this context.

\vskip 0.3cm
\noindent $\bullet$ {\bf Is the glassy state in some manganites of the same variety as
a standard spin-glass, or does it belong to 
a new class of ``phase-separated glasses''?}
The issue is subtle since, as of today, proper
definitions of glasses and associated transitions are still under much 
discussion. Can glasses be classified into different groups? Are 
glassy manganites a new class? What is the actual nature of the
``cluster glasses'' frequently mentioned in manganites? Important
work by Schiffer's group, by Levy, Parisi and collaborators
in Buenos Aires, and by many other researchers, has added key 
information to this subarea of manganite investigations. Studies of
time-dependent
phenomena in the $x$=0.5 regime
has provided fascinating results, that deserve further research.

Related with the previous item,
should nanocluster phase-separated states be considered as a
``new state of matter'' in any respect? This issue is obviously very important. In
order to arrive at an answer, the origin of the nanocluster formation 
must be fully clarified. Is it purely electronic-driven, or a
first-order transition rendered continuous by quenched
disorder, or strain-driven? 
If the inevitable disorder related to tolerance factors could
be tuned, what happens with the nanoclusters and critical temperatures?

\vskip 0.3cm
\noindent $\bullet$ {\bf What is the nature of the 
charge-ordered states at $x$$<$0.5, such as 
those in PCMO?}
It is accepted that these states ``resemble'' the CE-state in their charge
distribution, but what is the actual arrangement of spins and orbitals? Is the
excess of electronic charge distributed randomly in the $x$=0.5 CE
structure or is it uniformly distributed, for instance
by increasing uniformly the amount of charge  in the Mn$^{4+}$ sites? 
Theoretical studies are difficult at these hole concentrations. It is 
important to know whether entropy is large in the CO states at these
densities, to justify thermodynamically their existence
(as discussed by Khomskii). Otherwise,
how can a putative low-entropy CO state be stabilized at high temperature?
The opposite, a FM state stable above the CO state, is more
reasonable and has been already shown to be the case in simulations
by Aliaga and collaborators.

\vskip 0.3cm
\noindent $\bullet$ {\bf Is there any compound with 
a truly spin-canted homogeneous ground state?}
As far as I know, theoretical studies using robust techniques
have not been able to find spin-canted homogeneous
states in reasonable models for manganites (of course, if no
magnetic field is added). Are there experiments suggesting
otherwise? Thus far, experimental evidence for canting can be always
alternatively
explained through inhomogeneities in the ground state. A counterexample
are perhaps bilayers in the direction perpendicular to the planes
(see the many results by the Argonne group), but this may
be a different kind of state, unrelated to the original proposal by DeGennes 
that postulated a homogeneous spin-canted state interpolating 
between FM and AF limits.

\vskip 0.3cm
\noindent $\bullet$ {\bf The temperature dependence of the
d.c. resistivity of manganites has not
been sufficiently analyzed}  Can non-Fermi-liquid behavior be shown
to be present in metallic manganites, as it occurs in many other
exotic metals?

\vskip 0.3cm
\noindent $\bullet$ {\bf Are the nanoclusters found 
in manganites and cuprates (see below)
characteristic of other oxides as well?} 
The answer seems to be yes. In Chapter 21 of \cite{book},
many materials that behave similarly to Mn- and Cu-oxides are listed.
Nickelates are another family of compounds that have 
stripes and charge-ordering
competing with antiferromagnetism. 
Below, other materials with similar characteristics are briefly discussed.
These analogies are more than accidents. They suggest that many
compounds are intrinsically inhomogeneous. Theories based on homogeneous
states appear unrealistic.

\vskip 0.3cm
\noindent $\bullet$ {\bf Are Eu-based 
semiconductors truly described by ferromagnetic polarons
as believed until recently,
or is a nanocluster picture more appropriate?} 
The recent
studies  by Lance Cooper's group using Raman scattering suggest
the existence of close analogies between Eu semiconductors and manganites.
Perhaps phase separation dominates in Eu compounds as well, and the
long-held view of Eu-semiconductors as containing simple 
``ferromagnetic polarons'' (one carrier, with a spin polarized cloud around)
should be revised.

\section{\bf Some of the Unsolved Theoretical Issues}

\noindent $\bullet$ {\bf The study of models for manganites is far 
from over!} Although
much progress has been made  \cite{book,review}
many important issues are still unexplored
or under discussion. For example, the phase diagram in {\it three
dimensions} of the two-orbitals model may contain many surprises.
It is already known that the 1D  and 2D models 
have a rich phase diagram, with a variety of competing phases.
In addition, studies by Hotta et al.
in two dimensions including cooperative effects
have also revealed the presence of stripes 
at densities $x$=1/3 and 1/4, and probably others. Then, one can
easily 
imagine a surprisingly rich phase diagram for bilayers or 3D systems.
Perhaps the dominant phases will still be 
the A-type AF at $x$=0, ferro-metal at $x$$\sim$0.3,
CE-type at $x$=0.5, C-type at $x$$\sim$0.75, and G-type at $x$=1.0. However,
the details remain unclear.

\vskip 0.3cm
\noindent $\bullet$ {\bf 
The theory of phase separation should be made more quantitative.} 
Can a rough temperature dependence of
the d.c. resistivity be calculated within the percolative
scenario? Certainly we already have resistor-network calculations
that match the experiments, but not a simple anybody-can-use formula.
This is a complicated task due to the difficulty in handling 
inhomogeneities.

\vskip 0.3cm
\noindent $\bullet$ {\bf Is the ``small'' $J_{\rm AF}$
 coupling between localized $t_{2g}$ spins 
truly an important coupling for manganites?} 
The Heisenberg coupling between localized spins appears to play a key
role in the stabilization of the A-type AF state at $x$=0. This small
coupling
selects whether the system is in a  FM, A-, C- or G-type state, namely
its influence is amplified in the presence of nearly degenerate states.
This coupling is also important in
the stabilization of the correct spin arrangement for the CE-phase (if
$J_{\rm AF}$ is too small, ferromagnetism wins, 
if too large G-type antiferromagnetism wins),
and in the charge stacking of the CE-phase, according to studies by
Yunoki, Hotta, Terakura, Khomskii and others.
This key role
appears not only in Jahn-Teller but also in 
Coulombic-based theories as well, as shown by Ishihara, Maekawa and
collaborators. Are there alternative
mechanisms for stabilization of A-, CE- and charge stacking
states? The importance of $J_{\rm AF}$ may be magnified by the 
competition between
many phases in manganites, with several nearly degenerate states.

\vskip 0.3cm
\noindent $\bullet$ {\bf Why is the CE-state 
so sensitive to Cr-doping?} Experimentally it is not expected
that a robust charge-ordered state could be destabilized by
a relatively very small percentage of impurities. Can this be
reproduced in MC simulations?  Note that high-Tc cuprates are
also very sensitive to impurities. I believe that 
materials near percolative transitions are naturally
sensitive to disorder. These effects may provide even more evidence for
the relevance of inhomogeneous states in oxides.

\vskip 0.3cm
\noindent $\bullet$ {\bf For the explanation of CMR, 
is there a fundamental difference
between JT- and Coulomb-based theories?} Technically, it is quite
hard to handle models where simultaneously the Coulomb Hubbard 
interactions as well as the electron-phonon couplings are large.
However, thus far for CMR 
phenomena the origin (JT vs. Coulomb) of the competing phases does not
appear to be crucial, but the competition itself is. Is this correct?

\vskip 0.3cm
\noindent $\bullet$ {\bf In the calculations 
by Burgy et al. \cite{burgy}, phenomenological models
were used for CMR. Can a large MR effect be obtained with more realistic
models?} Of course the calculations would be very complicated in this 
context, if unbiased robust many-body techniques are used, due to
cluster size limitations.

\vskip 0.3cm
\noindent $\bullet$ {\bf Are 
there models with spin canted homogeneous ground states?}
Thus far, when models for manganites were seriously studied with unbiased
techniques, no homogeneous spin canted states have been identified (at
zero magnetic field). Perhaps other models?

\vskip 0.3cm
\noindent $\bullet$ {\bf Can a model develop a charge-ordered AF phase at
intermediate temperatures while having a ferromagnetic metallic phase at
low temperatures?} This is quite hard and perhaps can only occur if the
CO phase has an associated high entropy, as has been discussed 
by Khomskii (see similar discussion in the previous section).

\section{\bf Are There Two Types of CMR?}

\begin{figure}[h]
\includegraphics[width=.55\textwidth]{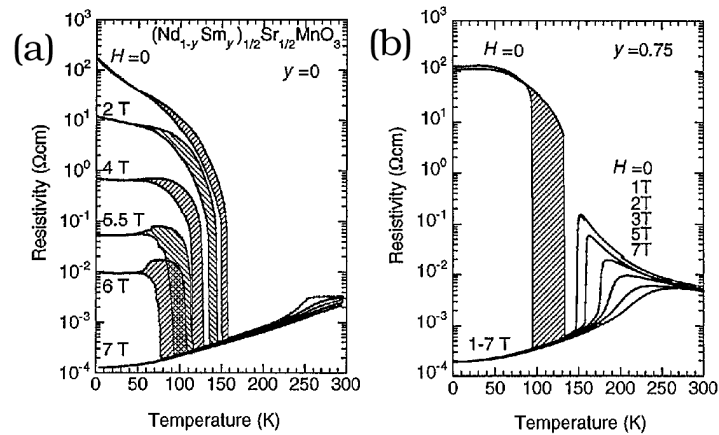}
\caption[]{
Temperature dependence of resistivity under various magnetic fields for 
 $\rm (Nd_{1-{\it y}} Sm_{\it y})_{1/2} Sr_{1/2} Mn O_3$ with $y$=0 (a) and 0.75 (b). The
 hatched area represents thermal hysteresis. Results
 from Tokura {\it et al.} \cite{tokura96}.
\label{twotypes.eps}
}
\end{figure}

The manganite $\rm (Nd_{1-{\it y}} Sm_{\it y})_{1/2} Sr_{1/2} Mn O_3$ investigated 
by Tokura and collaborators has an interesting behavior, shown in
Fig. \ref{twotypes.eps}.
This compound presents {\it two} types of
CMR phenomena: (1) At temperatures in the vicinity of 250~K ($y$=0) and
150~K ($y$=0.75) a somewhat
``standard'' CMR is observed. Here, by standard it is understood the CMR
behavior described in reviews and \cite{book}, with the canonical shape
of the resistivity vs. temperature.
However, (2) at lower temperatures where the system is insulating, a
{\it huge} MR effect is observed as well. Results for two values of ``$y$'' are
shown in the figure. Is this indicative of two independent mechanisms
for CMR?

Related with this issue is the recent work of Fernandez-Baca and
collaborators studying $\rm Pr_{0.70} Ca_{0.30} Mn O_3$, where
they reported the discontinuous character of the insulator to metal
transition induced by an external field. Those authors argued correctly 
that the CMR phenomenon can be more complex than the 
percolation of FM clusters proposed for the standard CMR.
Their results are in qualitative agreement with those of Fig.\ref{twotypes.eps}
where, at low temperatures, a fairly abrupt transition is observed.
In fact, it is known that even in percolative processes such as those
in LPCMO (see work of Cheong and collaborators), 
hysteresis has been found in the resistivity. This suggests that
a {\it mixture} of percolation with first-order features
could be at work in manganites, as theoretically discussed by Burgy 
{\it et al.} \cite{burgy,book}.

\begin{figure}[h]
\includegraphics[width=.37\textwidth]{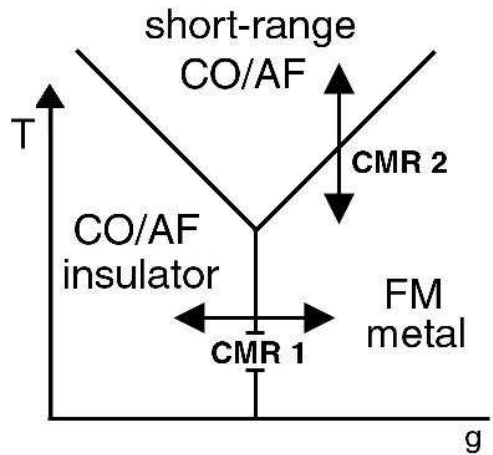}
\caption[]{
Schematic representation of the generic phase diagram in the presence of
competing metal and insulator, and for quenched disorder not 
sufficiently strong to destroy entirely the first-order transition at
low temperatures.
$g$ is a generic variable needed to transfer the system from one
phase to the other. CMR1 and CMR2 are the regions with two types of large
MR transitions, as described in the text and in \cite{book}.
\label{twocmr.eps}
}
\end{figure}

Can theory explain the presence of two types of CMR transitions? The
answer is tentatively yes, although more work is needed to confirm the picture.
The main reason for expecting two types of CMR 
is already contained in the phase diagram
of the models studied by Burgy et al. \cite{burgy} to address the
competition of two phases, as schematically reproduced in
Fig.\ref{twocmr.eps}. There, two possible regions with CMRs effects 
are shown. One corresponds
to the ``standard'' region, at the transition from a ``clustered'' 
short-range ordered
phase to the FM metallic phase with decreasing temperature (CMR2).
However, at low temperatures,
and if the quenched disorder is not too strong, 
the transition between the competing phases can remain of first
order in a finite range of temperatures (see Fig.1 and discussion in Sec.II).
As a consequence, if the system is very close to the metal-insulator
transition and on the insulating side, 
a small magnetic field can cause a first-order transition
between the two phases (in the CMR1 region in the figure), which will 
imply a dramatic and discontinuous
change in the resistivity. This effect has been recently confirmed
in simulations of the two-orbital model with Jahn-Teller phonons by
Aliaga and collaborators (to be published). Clearly, the theory predicts
two types of CMR transitions and this is already in agreement with
experiments. Note that if there were an external means to favor the
CO/AF phase over the FM phase (for instance by an ``external staggered 
field''), then the process could be reversed and a metal to insulator
exotic transition would be induced. See more about this when potential colossal
effects are discussed for cuprates.

\section{EVEN MORE GENERAL OPEN QUESTIONS}

At the risk of sounding naive, here are very general issues that 
in my opinion should also be addressed by experts:

\vskip 0.5cm

\begin{figure}[h]
\includegraphics[width=.30\textwidth]{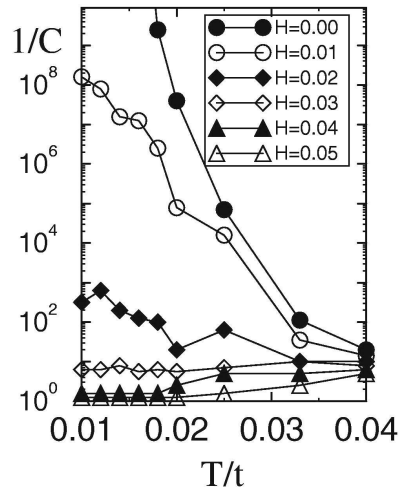}
\caption[]{
Inverse conductivity of the half-doped one-orbital model on a 64-site
chain at several values of the magnetic field $H$ (in units of the hopping) 
as indicated. Note the enormous changes
in the conductivity with increasing fields. Couplings and details
can be found in the original work by Mayr and collaborators \cite{mayrPRL}.
\label{eps15.35}
}
\end{figure}

{\bf Can quasi one-dimensional manganites be prepared experimentally?}
Perhaps chemists may already know that this is impossible, but here
let us just say that in the area of cuprates it was quite instructive 
to synthesize  copper-oxides with {\it chain or ladder structures}. These
quasi-1D systems can usually be studied fairly accurately by theorists,
and concrete quantitative predictions can be made, both for static
as well as dynamical properties. Regarding the CMR effect,
there are already calculations (see Fig.\ref{eps15.35}) that show a
very large MR effect in 1D models. In a 1D system
a single perfectly antiferromagnetic
link is sufficient to block the movement of charge (since
its effective hopping in the large Hund coupling limit 
is zero), creating a huge resistance. Small fields can slightly bend those
AF oriented spins, allowing for charge movement and decreasing 
rapidly the resistivity
by several orders of magnitude, as shown in the figure.
 
\vskip 0.3cm
{\bf Should we totally exclude superconductivity as a possibility  in manganites?}
Naively, the presence of superconductivity in 
transition-metal oxides of the $n$=3 shell 
should not be necessarily restricted to just the 
cuprates. In the Cu-oxide context, superconductivity appears when the
insulator is rendered unstable by hole doping. Antiferromagnetic 
correlations are among the leading candidates to explain the $d$-wave
character of the superconducting state. In the high-doping side
of the superconducting region, a Fermi liquid exists even at low 
temperature. Is there a doped-manganite compound, 
likely a large bandwidth one, that
does not order at low temperatures and maintains a metallic character?
Or, alternatively, can a metallic manganite compound
 become paramagnetic upon the application
of, e.g., high pressure, at low temperatures?
Searching for metal-insulator transitions near these
{\it quantum critical points}, if they exist, may
lead to surprises. If it is confirmed that, even in these circumstances,
no superconductivity is found, then the $S$=1/2 spins of cuprates (as
opposed to the higher spins of other materials) is likely to play
a key role in the process. Studies  by Riera and collaborators
\cite{riera} have shown that the prominent
presence of ferromagnetism and phase
separation as the spin grows, may render superconductivity unstable
in Mn oxides.

\vskip 0.3cm
{\bf Technical applications of manganites remain a possibility.}
There are two areas of technologically-motivated investigations. One
is based on materials with high $T_{\rm C}$, at or above room temperature,
which may be useful as nearly half-metals 
in the construction of multilayers spin-valve-like devices. The group
of Fert has made progress in this area recently, and there are several
other groups working on the subject.
Another area is the investigations
of thin-films, exploiting CMR as an intrinsic property of these materials. 
Especial treatments of those films 
may still lead to a large MR at high temperatures and small fields, as
needed for applications.
Of course, there is a long way before CMR even matches  the remarkable
performance of the giant-MR (GMR) devices.

\section{Is the Tendency to Nanocluster Formation Present in Other Materials?}

We end this informal 
paper with a description of materials that present properties
similar to those of the Mn-oxides, particularly regarding inhomogeneities, phase
competition, and the occurrence 
of a $T^*$ (for a detailed
long list of materials see \cite{book}).
It is interesting to speculate that all of
these compounds share a similar phenomenology. Then, by investigating
one particular 
system progress could be made in understanding the others as well. 
We start with the famous high-T$_c$ compounds, continue with diluted
magnetic semiconductors -- of much interest these days due to spintronic
applications -- and finish with
organic and heavy-fermion materials.

\subsection{Phase Separation in Cuprates}

\begin{figure}[h]
\includegraphics[width=.3\textwidth]{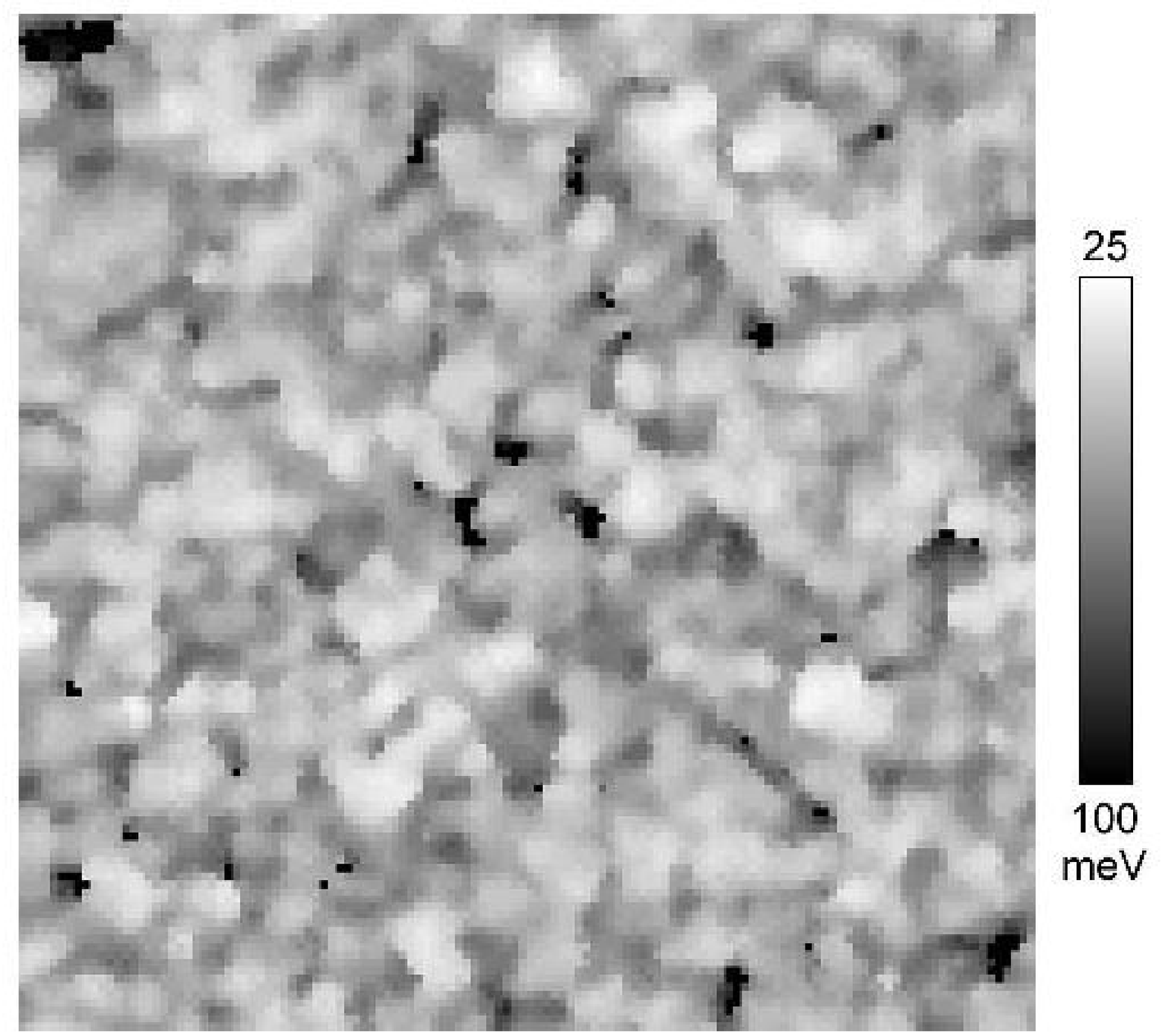}
\includegraphics[width=.3\textwidth]{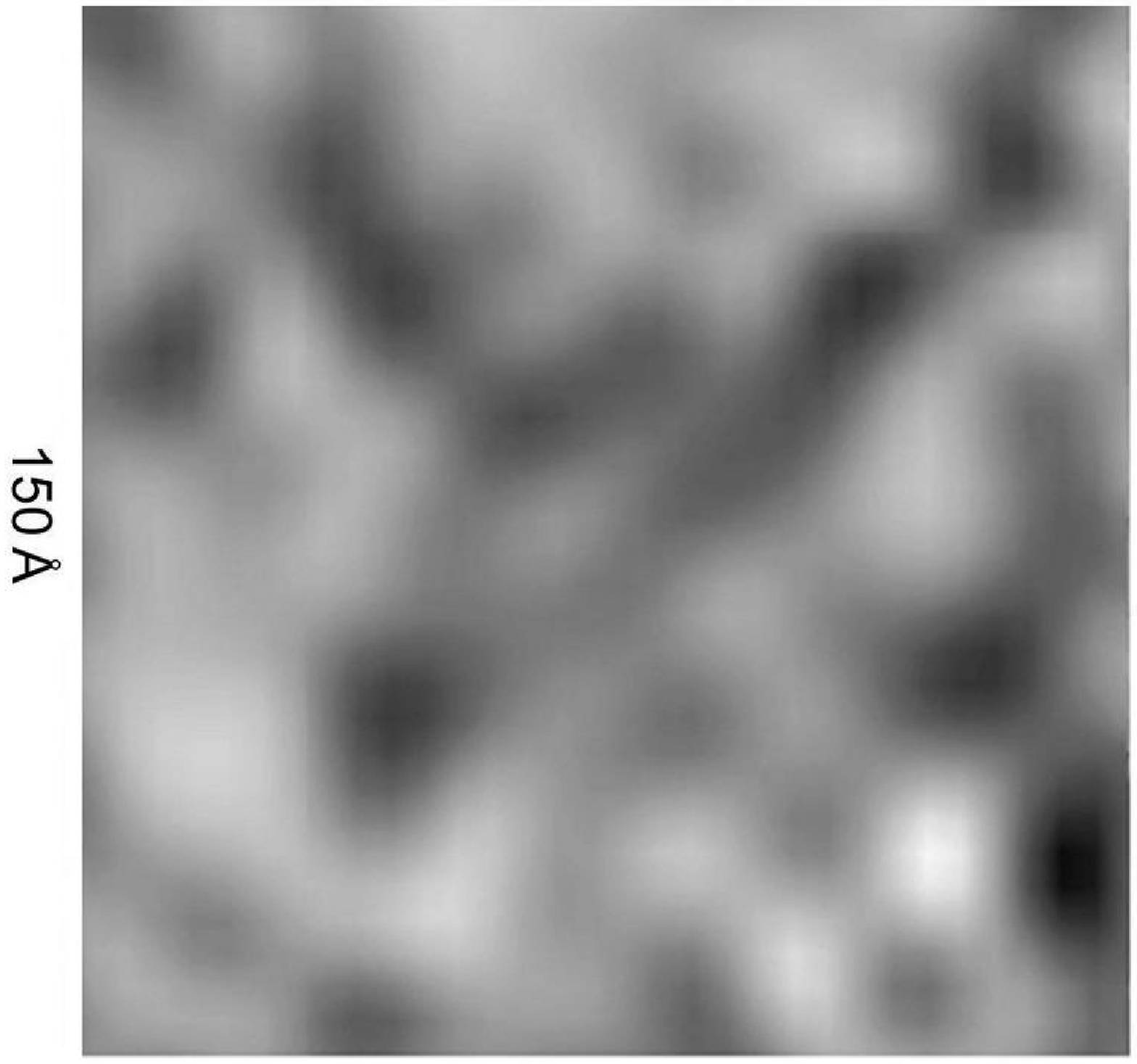}
\caption[]{ Spatial variation of the superconducting energy gap
of Bi2212 as reported by Pan {\it et al.} \cite{pan}, using STM
techniques. Shown on the left is a 600$\times$600 $\rm \AA$ area. 
On the right a 150$\times$150 $\rm \AA$ subset is enlarged. The
gap scales are also shown.
\label{pan2.eps}
}
\end{figure}

Considerable work has been recently devoted to the study of
inhomogeneous states in cuprates. In this context, the issue of stripes as
a form of inhomogeneity was raised several years ago  after experimental
work by Tranquada and collaborators and theoretical work by
Kivelson, Zaanen, Poilblanc, and others.
However, recent results obtained with STM techniques by Davis, Pan, and
collaborators have
revealed inhomogeneous states of a more complex nature. They
appear in the surface of one of the most studied superconductors
Bi2212 (but they may be representative 
of the bulk as well), at low temperatures in the superconducting
phase, both in the underdoped and optimally doped regions.
Figure \ref{pan2.eps} shows the spatial distribution of 
$d$-wave superconducting
gaps, where the inhomogeneities are notorious.
Indications of similar
inhomogeneities have been observed using a variety of 
other techniques as well
(the list of relevant experiments 
is simply too large to be reproduced here. 
For a partial list see \cite{book}). Currently, this is a
much debated area of research in high critical temperature 
superconductors, and it is unclear whether cuprates are intrinsically
inhomogeneous in all its forms, or whether the inhomogeneities are
a pathology of only a fraction of the Cu oxides. It is also unclear to
what extend the notion of stripes, with at least partial order in its
dynamical form, survives the Bi2212 evidence of
inhomogeneities where the patches are totally random.

Note that recent NMR work on YBCO by Bobroff et al. 
has not reported inhomogeneities,
suggesting that phase separation may not be a generic feature of the
cuprates. Similar conclusions were reached by Yeh and collaborators 
through the analysis of quasiparticle tunneling spectra.
Clearly, more work is needed to clarify the role of inhomogeneities in
the cuprate compounds.

\subsection{Colossal Effects in Cuprates?}

\begin{figure}[h]
\centering
\includegraphics[width=.65\textwidth]{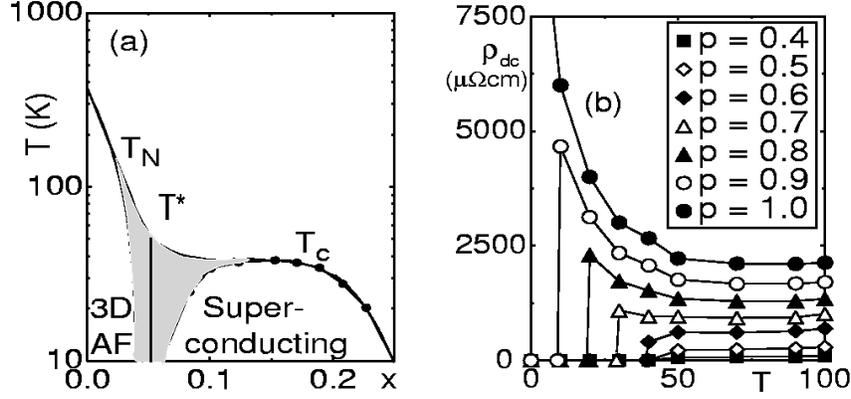}
\caption[]{
(a) Phase diagram of the 214 high temperature superconductor, as conjectured
by Burgy {\it et al.} \cite{burgy}. 
A possible first-order line (thick vertical line) separating the clean-limit
superconducting and antiferromagnetic insulating regimes in the low
hole-density region is shown. With
disorder, the grey region is generated, exhibiting coexisting clusters.
As in the manganite context discussion,
$T^*$ is the remnant of the clean-limit ordering temperature.
(b) Resistivity of the state conjectured in (a) using a crude resistor-network
approximation, as calculated by Mayr. 
More details can be found in the original reference
\cite{burgy}. $p$ is insulating fraction at 100~K. These
fractions are also temperature dependent. Figure (b) simply illustrates
the possibility of a percolative transition in the cuprate context.
\label{eps.prl2001.4}
}
\end{figure}

The results discussing the general aspects
of the competition between two ordered phases in
the presence of quenched disorder (see for instance the phase
diagram in Sec.II) can in principle be applied to the cuprates as well,
where superconductivity competes with an antiferromagnetic state 
(the latter perhaps including stripes).
In fact, the phase diagram of the single-layer 
high-$T_{\rm c}$ compound
$\rm La_{2-{\it x}} Sr_{\it x} Cu O_4$ contains a regime widely known as the
``spin glass'' region. In view of the results in \cite{book,burgy},
it is conceivable that this spin-glass
region could have emerged, due to the influence of disorder, from a clean-limit
first-order phase transition separating the doped antiferromagnetic
state (again, probably with stripes) and the superconducting state. The
resulting phase diagram would be as in Fig. \ref{eps.prl2001.4}(a). The
grey region exhibits a possible coexistence of small locally-ordered clusters, 
globally forming a disordered phase. If this conjecture is correct, then 
{\it small superconducting islands should exist in the spin-glass regime},
an effect that may have already been observed by Iguchi {\it et al.} 
\cite{iguchi}. The results of Ong et al. on the Nernst effect may also find
an explanation with this scenario.

If preformed superconducting regions are indeed 
present in underdoped
cuprates, then the ``alignment'' of their order parameter (i.e.
alignment of the local phases $\phi$ that appear
in the complex-number order
parameter $\Delta$=$|\Delta|$$e^{i\phi}$ for different islands)
should be possible upon the influence of small external perturbations.
This is similar to the presence
of preformed ferromagnetic clusters in manganites, which align their
moments when a
relatively small external magnetic field is applied (see \cite{burgy}
and also Cheong and collaborators). 
Could it be that 
``colossal'' effects occur in cuprates and other materials
as well? This is an
intriguing possibility raised by Burgy {\it et al.} \cite{burgy}. 
There are already experiments 
by Decca and collaborators \cite{decca} that
have reported an anomalously large ``proximity effect'' in 
underdoped $\rm Y Ba_2 Cu_3 O_{6+\delta}$. In fact, Decca {\it et al.}
referred to the effect 
as ``colossal proximity effect''. Perhaps colossal-type
effects are more frequent than previously expected, and they are
prominent in manganites simply because one of the competing phases is
a ferromagnet which can
be favored by an easily available
uniform magnetic field. If a simple external
perturbation would favor charge-ordered, antiferromagnetic, or
superconducting states, large effects may be observed 
in the region of
competition with other phases. If this scenario is correct, several
problems should be percolative, and studies by Mayr in
\cite{burgy} suggest that a
resistor network mixing superconducting and insulating islands 
could roughly mimic results for cuprates (see Fig.9(b)). 
However, obviously more work is needed to test 
these very sketchy challenging ideas. 
Note that this scenario for cuprates is totally
different from the ``preformed Cooper pairs''  ideas of Randeria and collaborators
where the state above the critical temperature is homogeneous, but contains
pairs that have not condensed. Here, the superconducting regions can be fairly
large, containing probably hundreds of pairs, 
and are expected to move very slowly with time.

Even at a phenomenological level, the order of the transition between
an antiferromagnet and a superconductor is still 
unknown. Mayr and collaborators
are carrying simulations of toy models with both phases which hopefully
will clarify these matters. The possibilities for the phase diagrams of two
competing phases --according to general
argumentations based on Landau free energies-- 
involve bi-, tri-, and tetra-critical order. In the context
of manganite models, 
the first two appear to be favored for the AF-FM competition
since first-order transition at low temperatures have been found. 
However, in cuprates the tetracritical possibility in the clean limit
has been proposed by Sachdev and collaborators as well. 
This would lead to local
coexistence of antiferromagnetism and $d$-wave superconductivity in the
underdoped region. Some experiments support this view \cite{sachdev}.
Recent theoretical work also favors the emergence of tetracriticality from
this competition \cite{sachdev2}, although a first-order transition is 
not excluded.
More work is needed to clarify this elementary aspect of the problem.

The results for manganites and cuprates
have also similarities with the physics
of superconductivity in granular and highly disordered metals
\cite{strongin}. In this 
context, a critical temperature is obtained when the phases of the order
parameters in different grains locks. This temperature is much smaller
than the critical temperature of the homogeneous system. In manganites,
the Curie or charge-ordering temperature may also be much higher in a
clean system than in the real system, as discussed here 
and in \cite{book,review}. 

\subsection{Relaxor Ferroelectrics and $T^*$}

The so-called {\it relaxor ferroelectrics} are an interesting family of
compounds, with ferroelectric properties and diffuse phase
transitions \cite{cross}. The analogies with manganites are
notorious. Representative cubic
materials, such as $\rm Pb Mg_{1/3} Nb_{2/3} O_3$, exhibit a glass-like
phase transition at a freezing temperature $T_{\rm g}$$\sim$230~K. Besides the
usual features of a glassy transition, a broad frequency dependent peak
in the dielectric constant has been found. The transition does not involve
long-range ferroelectric order, providing a difference with most
manganites that tend to have some form of long-range order at low
temperature (although there are several with glassy behavior). 
However, a remarkable similarity is the appearance of
another temperature scale, the so-called Burns temperature \cite{burns}, 
which is $\sim$650~K in $\rm Pb Mg_{1/3} Nb_{2/3} O_3$. 
Below this temperature, polar
{\it nanoregions} are formed. The Burns temperature appears to be the analog
of the $T^*$ temperature of manganites discussed here.
These analogies between ferroelectrics and Mn-oxides
should be explored further.
For example, is there some
sort of ``colossal'' effect in the relaxor ferroelectrics? Are there two
or more phases in competition? Can we alter the chemical compositions
and obtain phase diagrams as rich as in manganites, including regimes with
long-range order?

\subsection{Diluted Magnetic Semiconductors and $T^*$}

\begin{figure}[h]
\includegraphics[width=.25\textwidth]{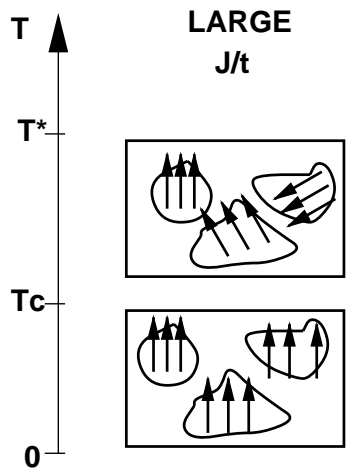}
\caption[]{
Schematic representation of the clustered state proposed for
diluted magnetic semiconductors in \cite{alvarez}.
At $T^*$, uncorrelated
ferromagnetic clusters are formed.
At $T_{\rm C}$, they orient their moments in the same direction.
\label{diluted.eps}
}
\end{figure}

Diluted magnetic semiconductors (DMS) based on III-V compounds
are recently attracting considerable attention due to their combination
of magnetic and semiconducting properties, that may lead to
spintronic applications \cite{ohno2}. 
Ga$_{1-x}$Mn$_x$As is the most studied of these compounds with
a maximum Curie temperature $T_{\rm C}$$\approx$$110$K at low doping $x$,
and with a carrier concentration $p$=0.1$x$ \cite{ohno2}.
The ferromagnetism is carrier-induced,
with holes introduced by doping mediating the interaction between
$S$=5/2 Mn$^{2+}$-spins. Models similar to those proposed for manganites,
with coupled spins and carriers, have been used for these compounds.
Recently, the presence of a $T^*$ has also been
proposed  by Mayr {\it et al.} 
and Alvarez {\it et al.} \cite{alvarez}, using Monte Carlo methods
quite similar to those discussed in the manganite context.
The idea is that in the random
distribution of Mn spins, some of them spontaneously
form clusters (i.e., they lie
close to each other). These clusters can magnetically order by the
standard Zener mechanism at some temperature,
but different clusters may not correlate
with each other until a much lower temperature is reached. This
situation is illustrated in Fig.\ref{diluted.eps}. Naively, we
expect that above the true $T_{\rm C}$, the clustered state
will lead to an insulating behavior in the resistivity, as it happens
in Mn-oxides (recent work by Alvarez has confirmed this hypothesis). 
This, together with the Zener character of the ferromagnetism,
unveils unexpected similarities between DMS and Mn-oxides that should be
studied in more detail.

\subsection{Cobaltites}

Recent work has shown that cobaltites, such as $\rm La_{1-{\it x}} Sr_{\it x} 
Co O_3$, also present tendencies toward magnetic 
phase separation \cite{cobaltites}. 
NMR studies have established the coexistence of ferromagnetic regions,
spin-glass regions, and hole-poor low-spin regions in these
materials \cite{cobal2}. This occurs at all values $x$ from 0.1 to 0.5.
There are interesting similarities between these results and the
NMR results of Papavassiliou and collaborators for the manganites,
that showed clearly the tendencies toward mixed-phase states. The
analysis of common features (and differences) 
between manganites and cobaltites should
be pursued in the near future.

\subsection{Organic Superconductors, Heavy Fermions, and $T^*$}

\begin{figure}[h]
\centering
\includegraphics[width=.7\textwidth]{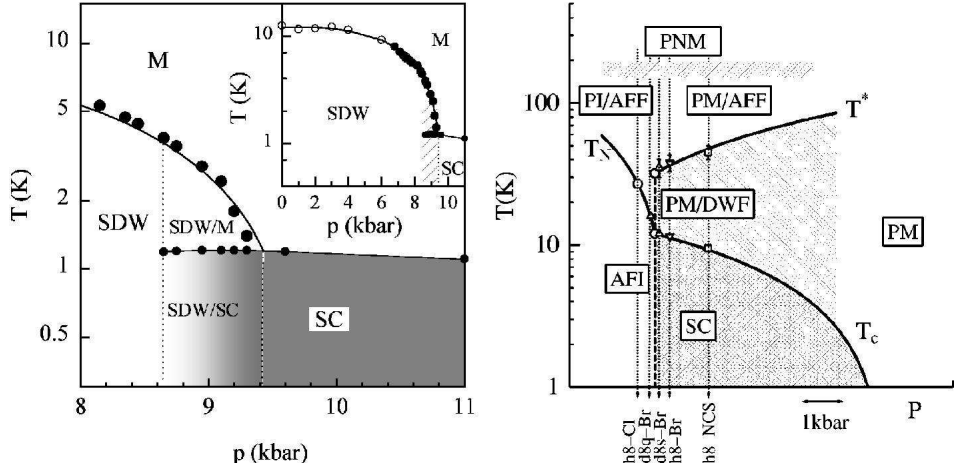}
\caption[]{
{\it (left)} Phase diagram of (TMTST)$_2$PF$_6$  \cite{vuletic}. SDW, M,
and SC indicate spin-density-wave, metallic, and superconducting
regimes. The gradient in shading indicates the amount of
SC phase. 
{\it (right)} Phase diagram of $\kappa$-(ET)$_2$X \cite{sasaki}. 
PNM means paramagnetic non-metallic, AFF
a metallic phase with large AF fluctuations, and DWF is a density-wave with
fluctuations. Note the first-order SC-AF transition (dashed lines),
as well as the presence of a $T^*$ scale. Other details 
can be found in \cite{sasaki}.
\label{organic.eps}
}
\end{figure}

There are other families of materials that also present a competition
between superconductivity and antiferromagnetism, as the cuprates do. 
One of them is
the group of organic superconductors \cite{organic-book}. 
The large field of organic superconductivity will certainly not be
reviewed here, but some recent references will be provided to
guide the manganite/cuprate experts into this interesting area of research. In
Fig.\ref{organic.eps} (left), the phase diagram \cite{vuletic}
of a much studied  
material known as (TMTST)$_2$PF$_6$ is shown. 
In a narrow region of pressures, a
mixture of SDW (spin-density-wave) and SC (superconductivity) is
observed at low temperatures. This region may result from coexisting
domains of both phases, as in the FM-CO competition in manganites.
The two competing phases have the same electronic density and the
domains can be large. Alternatively, the coexistence of the two order
parameters can be local, as in tetracritical phase diagrams.
The layered organic superconductor $\kappa$-(ET)$_2$Cu[N(CN)$_2$]Cl 
has also a phase diagram with coexisting SC and AF \cite{lefebvre}.
In addition, other materials of the same family present a $first$-$order$
SC-AF transition  at low temperatures, according to the phase diagram
recently reported in \cite{sasaki}. This result
is reproduced in Fig.\ref{organic.eps} (right), where a characteristic
scale $T^*$ produced by charge fluctuations is also shown. The
similarities with results for manganites are strong:
first-order transitions or phase coexistence, and the presence of a $T^*$. 
It remains to be investigated whether
the similarities are accidental
or reveal the same phenomenology observed in competing phases of 
manganites.


\begin{figure}[h]
\includegraphics[width=.43\textwidth]{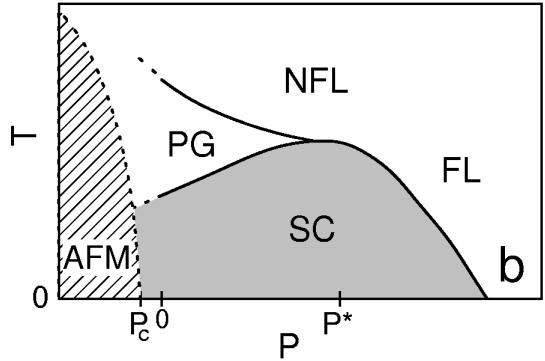}
\caption[]{
Schematic temperature-pressure phase diagram of some Ce-based heavy
fermions, from \cite{sidorov}. The notation is standard.
\label{heavy.fermions.eps}
}
\end{figure}

Another family of materials with competing SC and AF are the heavy
fermions. 
As in previous cases, here just a few references are provided
to illustrate similarities with other compounds. 
In Fig.\ref{heavy.fermions.eps}, a schematic phase diagram of Ce-based
heavy fermions is shown
\cite{sidorov}. Note the presence of a pseudogap (PG) region,  with non
Fermi-liquid (NFL) behavior at higher temperatures. The PG phase is
reminiscent of the region between ordering temperatures and $T^*$,
discussed in manganites and cuprates, and it may arise from
phase competition. In addition, clusters in U- and Ce-alloys were discussed in
\cite{castroneto}, and spin-glass behavior in $\rm CeNi_{1-{\it x}}Cu_{\it x}$
was reported in \cite{garcia-ce}. Recently, Takimoto {\it et al.}
\cite{takimoto} found interesting formal analogies between models for
manganites and for $f$-electron systems that deserve further studies.

\section{Conclusions}

In recent years a large effort has been devoted to the study of
manganese oxides. Considerable progress has been achieved. Among the
most important unveiled aspects is the presence of tendencies toward
inhomogeneous states, both in experiments and in simulations of models.
However, considerably more work is needed to fully understand these
challenging compounds and confirm the relevance of nanoclusters for the
CMR effect.

Considering a more global view of the problem, recent years have brought
a deeper understanding of the many analogies among a large fraction of 
the materials belonging to the correlated-electron family.
The analogies between organic, cuprate,
and heavy-fermion superconductors are strong. 
The SC-AF phase competition with the presence of a
$T^*$, pseudogaps, phase
coexistence, and first-order transitions, is formally similar to
analogous phenomena unveiled in the FM-CO competition for manganites.
In addition, all of these compounds share a similar phenomenology above
the critical temperature, with nanoclusters formed. This phenomenon
occurs even in
diluted magnetic semiconductors, relaxor ferroelectrics,
 and other families of
compounds.
{\it The self-organization of clustered structures in the ground state 
appears to be a characteristic of many interesting materials,
and work in this promising area of investigations is just starting.}
This ``complexity'' seems a common 
feature of correlated electron systems, particularly
in the regime of colossal effects where small changes in external parameters
leads to a drastic rearrangement of the ground state.

\section{Acknowledgments}

The author thanks S. L. Cooper, J. A. Fernandez-Baca, G. Alvarez,
and A. Moreo, for comments and help in the preparation of  
this manuscript. The work is supported by the Division of Materials 
Research of the National Science Foundation (grant DMR-0122523).


\begin{thebibliography}{99}

\bibitem{book}  E. Dagotto,
{\it Nanoscale Phase Separation and Colossal Magnetoresistance},
Springer-Verlag, Berlin, November 2002 (www.springer.de).

\bibitem{review} 
A. Moreo, S. Yunoki and E. Dagotto, Science {\bf 283}, 2034 (1999);
E. Dagotto, T. Hotta, and A. Moreo,
Phys. Reports {\bf 344}, 1 (2001);
A. Ramirez, J. Phys.: Cond. Matt. {\bf 9}, 8171 (1997);
J. Coey, M. Viret, and S. von Molnar, Adv. Phys. {\bf 48}, 167 (1999);
Y. Tokura and Y. Tomioka, J. of Mag. and
Mag. Materials {\bf 200}, 1 (1999);
M. R. Ibarra and J. M. De Teresa,
Materials Science Forum Vols. {\bf 302-303}, 125 (1999);
Trans Tech Publications, Switzerland;
C. Rao {\it et al.}, J. Phys.: Cond. Matt. {\bf 12}, R83 (2000);
A. Moreo, Journal of Electron Spectroscopy and Related Phenomena {\bf
117-118}, 251 (2001);
M. B. Salamon and M. Jaime, Rev. Mod. Phys. {\bf 73}, 583 (2001);
M. Imada, A. Fujimori, and Y. Tokura, Rev. Mod. Phys. {\bf 70}, 1039 
(1998); E. Dagotto, Rev. Mod. Phys. {\bf 66}, 763 (1994);
C. N. R. Rao and B. Raveau, editors, {\it Colossal Magnetoresistance,
Charge Ordering, and Related Properties of Manganese Oxides}, World
Scientific, Singapore, 1998;
T. A. Kaplan and S. D. Mahanti, editors,
{\it Physics of Manganites},
Kluwer Academic/Plenum Publishers, New York, 1998;
Y. Tokura, editor, 
{\it Colossal Magnetoresistive Oxides}, Gordon and Breach, 2000;
N. Mathur and P. Littlewood, Phys. Today, January 2003, page 25;
and many others.

\bibitem{newphases} {\it 
Unveiling New Magnetic Phases of Undoped and Doped Manganites},
T. Hotta, M. Moraghebi, A. Feiguin,
A. Moreo, S. Yunoki, and E. Dagotto, cond-mat/0211049.
Recently, a new FM charge-ordered phase was observed experimentally
J. Loudon {\it et al.}, Nature {\bf 420}, 797 (2002). This phase
has been predicted in \cite{newphase}

\bibitem{newnano} Very recent work V. Kiryukhin,  T. Y. Koo, H. 
Ishibashi, J. P. Hill, and S-W. Cheong, cond-mat/0301466, confirms
the essential role of the nanoscale inhomogeneities to understand
CMR physics. 

\bibitem{moreo} A. Moreo {\it et al.}, 
Phys. Rev. Lett. {\bf 84}, 5568 (2000).

\bibitem{burgy} J. Burgy, M. Mayr, V. Martin-Mayor, A. Moreo, E. Dagotto,
Phys. Rev. Lett. {\bf 87}, 277202 (2001).

\bibitem{slevin} K. Slevin, T. Ohtsuki, and T. Kawarabayashi,
Phys. Rev. Lett. {\bf 84}, 3915 (2000).

\bibitem{newphase} S. Yunoki {\it et al.}, 
Phys. Rev. Lett. {\bf 84}, 3714 (2000) and references therein.

\bibitem{tokura96} Y. Tokura, H. Kuwahara, Y. Moritomo, Y. Tomioka, and
A. Asamitsu, Phys. Rev. Lett. {\bf 76}, 3184 (1996).

\bibitem{mayrPRL} M. Mayr,
A. Moreo, Jose A. Verges, J. Arispe, A. Feiguin, and E. Dagotto,
Phys. Rev. Lett. {\bf 86}, 135 (2001).


\bibitem{riera}
J. Riera, K. Hallberg, and E. Dagotto, Phys. Rev. Lett. {\bf 79},
713 (1997).

\bibitem{pan} S. H. Pan, J. P. O'Neal, R. L. Badzey, C. Chamon, H. Ding,
J. R. Engelbrecht, Z. Wang, H. Eisaki, S. Uchida, A. K. Gupta, K-W. Ng,
E. W. Hudson, K. M. Lang, and J. C. Davis, Nature {\bf 413}, 282 (2001).

\bibitem{iguchi}
I. Iguchi, T. Yamaguchi, and A. Sugimoto, Nature {\bf 412}, 420 (2001).


\bibitem{decca} R. S. Decca, 
D. Drew, E. Osquiguil, B. Maiorov, and J. Guimpel,
Phys. Rev. Lett. {\bf 85}, 3708 (2000).

\bibitem{sachdev} See, for instance, B. Khaykovich {\it et al.},
Phys. Rev. B{\bf 66}, 014528 (2002); Y. S. Lee {\it et al.}, Phys. Rev.
B{\bf 60}, 3643 (1999).

\bibitem{sachdev2} P. Calabrese, A. Pelissetto, and E. Vicari,
cond-mat/0209580; and references therein.


\bibitem{strongin} M. Strongin, R. Thompson, O. Kammerer and J. Crow,
Phys. Rev. B{\bf 1}, 1078 (1970); Y. Imry and M. Strongin,
Phys. Rev. B{\bf 24}, 6353 (1981); and references therein. For related
more recent references see: A. F. Hebard and M. A. Paalanen,
Phys. Rev. Lett. {\bf 65}, 927 (1990); A. Yazdani and A. Kapitulnik,
Phys. Rev. Lett. {\bf 74}, 3037 (1995); M. Fisher, G. Grinstein, and
S. Girvin, Phys. Rev. Lett. {\bf 64}, 587 (1990); A. Goldman and
N. Markovic, Physics Today {\bf 51} (11), 39 (1998).

\bibitem{cross} For a review see L. E. Cross, Ferroelectrics {\bf 76}, 
241 (1987).

\bibitem{burns} G. Burns and B. A. Scott, Solid State Commun. {\bf 13},
423 (1973).


\bibitem{ohno2}  H. Ohno, Science, {\bf 281}, (1998);
T. Dietl, cond-mat/0201282.

\bibitem{alvarez} G. Alvarez, M. Mayr, and E. Dagotto,
Phys. Rev. Lett. {\bf 89}, 277202 (2002).
See also M. Mayr, G. Alvarez, and E. Dagotto, 
Phys. Rev. B{\bf 65}, 241202 (RC) (2002); and references therein.


\bibitem{cobaltites}
R. Caciuffo {\it et al.}, Phys. Rev. B{\bf 59}, 1068 (1999);


\bibitem{cobal2}
P. L. Kuhns, M. J. R. Hoch, W. G. Moulton, A. P. Reyes,
J. Wu, and C. Leighton, cond-mat/0301553, preprint.



\bibitem{vuletic} T. Vuleti\'c, P. Auban-Senzier, C. Pasquier, S. Tomi\'c,
D. J\'erome, M. H\'eritier, and K. Bechgaard,
cond-mat/0109031.

\bibitem{sasaki} T. Sasaki, N. Yoneyama, A. Matsuyama, and N. Kobayashi,
Phys. Rev. B{\bf 65} 060505 (2002).

\bibitem{organic-book} T. Ishiguro, K. Yamaji, and G. Saito,
{\it Organic Superconductors}, Springer-Verlag 1998. Phase
diagrams of relevance for our discussion can be found, e.g., in
Figures 3.14, 3.15, 3.19, 3.23, 4.14, 5.67, and 7.6.


\bibitem{lefebvre} S. Lefebvre {\it et al.}, Phys. Rev. Lett. {\bf 85},
5420 (2000).


\bibitem{sidorov} V. A. Sidorov, M. Nicklas, P. G. Pagliuso,
J. L. Sarrao, Y. Bang, A. V. Balatsky, and J. D. Thompson,
cond-mat/0202251; and references therein.

\bibitem{castroneto} A. H. Castro Neto and B. A. Jones,
Phys. Rev. B{\bf 62}, 14975 (2000).

\bibitem{garcia-ce} J. Garc\'ia Soldevilla {\it et al.},
Phys. Rev. B{\bf 61}, 6821 (2000).

\bibitem{takimoto} T. Takimoto, T. Hotta, T. Maehira, and K. Ueda,
cond-mat/0204023.

\end{thebibliography}
\end{document}